\documentclass[aps,prl,a4paper,reprint,showpacs,amsmath,amssymb,superscriptaddress,floatfix]{revtex4-1}

\usepackage{amssymb}
\usepackage{latexsym}
\usepackage[dvips]{graphicx}
\usepackage{epsfig}

\usepackage{dcolumn}
\usepackage{amsmath}
\usepackage{amsfonts}
\usepackage{bm}
\usepackage{color}

\newcommand{\be}{\begin{equation}}
\newcommand{\ee}{\end{equation}}
\newcommand{\bea}{\begin{eqnarray}}
\newcommand{\eea}{\end{eqnarray}}

\newcommand{\p}{\partial}
\newcommand{\s}{\sigma}

\newcommand{\la}{\langle}
\newcommand{\ra}{\rangle}

\newcommand{\ri}{\mbox{i}}

\renewcommand{\vec}[1]{{\bm #1}}

\begin{document}
\title{Chiral Spin Order in Kondo-Heisenberg systems}
\author{A. M. Tsvelik}
\affiliation{Condensed Matter Physics and Materials Science Division, Brookhaven National Laboratory, Upton, NY 11973-5000, USA}

\author{O. M. Yevtushenko}
\affiliation{Ludwig Maximilian University, Arnold Sommerfeld Center and Center for Nano-Science,
             Munich, DE-80333, Germany}

\date{\today }

\begin{abstract}
We demonstrate that Kondo-Heisenberg systems, consisting of itinerant electrons and localized
magnetic moments (Kondo impurities), can be used as a principally new platform to realize
scalar chiral spin order. The underlying physics is governed by a competition of the
Ruderman-Kittel-Kosuya-Yosida (RKKY) indirect exchange interaction between the local moments
with the direct Heisenberg one. When the direct exchange is weak and RKKY dominates the
isotropic system is in the disordered phase. A moderately large direct exchange leads to an
Ising-type phase transition to the phase with chiral spin order. Our finding paves the
way towards pioneering experimental realizations of the chiral spin liquid in low dimensional
systems with spontaneously broken time reversal symmetry. 
\end{abstract}

\pacs{
   75.30.Hx,   
   71.10.Pm,   
   72.15.Nj    
}

\maketitle

Interactions between magnetic moments usually lead to some kind of magnetic order where rotational
symmetry is broken and the order parameter is linear in spins
\cite{Auerbach}.  This is what happens in ferromagnets, antiferromagnets and all sorts of helimagnets.
Villain has demonstrated \cite{Villain-ScalChOrder} that, in addition to the magnetic order, helical magnets
possess a vector chiral order parameter. It is bilinear in spins and is related to the mutual orientation of
neighboring spins. This chiral order breaks the discrete symmetry and can exist even without the magnetic
order \cite{Pokrovsky-ChOrd}. The discovery of the vector chiral order has given rise to the idea that 
there could exist an order which includes a combination of three spins.
The corresponding  order parameter is a mixed product of three
neighboring spins, see $ {\cal O}_c $ in Eq.(\ref{ChOP}) below and Refs.\cite{Zee89,Baskaran89}.
It  breaks time-reversal and parity symmetries. Such a local order parameter is considered
as the key quantity for description of exotic magnetic phases \cite{Zee89}. In contemporary language,
$ {\cal O}_c $ is referred to as ``scalar chiral spin order'' and the state of matter with (spontaneously) 
broken time-reversal and parity symmetries but with conserved spin rotational symmetry is called Chiral 
Spin Liquid (CSL) \cite{SpLiqRev-2}. The seminal example possessing the CSL symmetry is the Kalmeyer-Laughlin
model \cite{CSL-FQHE-1,CSL-FQHE-2,CSL-3,CSL-4}. Its wave functions
demonstrate the topological behavior inherent in the fractional quantum Hall effect. 
Thus, the Kalmeyer-Laughlin model links spin liquids and topologically nontrivial states 
\cite{CSL-Marston,CSL-Girvin,CSL-Haldane,CSL-Cirac,CSL-Thomale1,CSL-Thomale2} and
can be called ``topological CSL''. An increasing interest in the topological CSL
\cite{CSL-Gurarie,CSL-Trebst,CSL-Neupert,CSL-Fradkin,CSL-Hickey,CSL-Penc,CSL-Tsv}
has been stimulated, in part, by a search for exotic (anyon) superconductivity \cite{Laughlin-SupCond,Chen-SupCond}
and by the physics of skyrmions \cite{Skyrm-1,Skyrm-2,Skyrm-3,Skyrm-4}. The latter can be realized in magnets
with the chirality resulting either from the lattice structure or from the Dzyaloshinskii-Moriya interaction 
\cite{Ch-DM,Ch-Interf,Ch-Latt,Wang_CSL_2017}.

Although the concept of CSL and its order parameter $ {\cal O}_c $ were introduced in  the 80-ties,  it still remains
unclear whether such a state can exist in realistic systems where time-reversal symmetry is not explicitly broken.
Numerous theoretical suggestions include spin systems with a complicated set of  either Heisenberg exchange interactions
extended far beyond nearest neighbors \cite{CSL-Balents,Chen14,Kagome-Balents} or multi-spin interactions
\cite{CSL-Thomale1,CSL-Thomale2}, Moat-Band lattices \cite{CSL-Moat}, and even laser-driven Mott insulators
\cite{CSL-Mott}. This list can be continued but, to the best of our knowledge, the question is still open and a reliable
experimental evidence of CSL governed by the spontaneously broken time reversal symmetry is still absent.

\begin{figure}[t]
   \includegraphics[width=0.42 \textwidth]{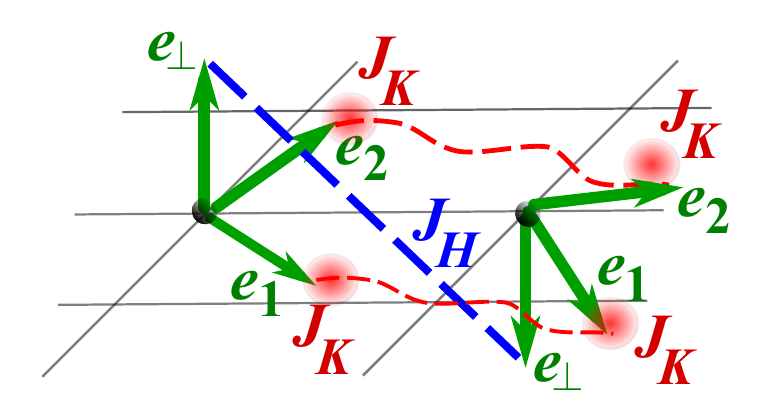}
   \caption{
        (color on-line) Competition between two different spin interactions
        in KHS: The spin on each lattice site is decomposed in terms of an orthonormal
        triad $ \vec{e}_{1,2,3} $ (green arrows) with $ \vec{e}_\perp = (-1)^{N(\vec{r})} \vec{e}_3 $,
        see Eq.(\ref{spins}). The RKKY exchange
        interaction (red dashed lines) is mediated by electrons (red circles) and favours
        helical-like configuration of the vectors $ \vec{e}_{1,2} $.
        The Heisenberg exchange interaction (blue dashed line) favours antiparallel orientation
        of $ \vec{e}_{\perp} $ on neighboring lattice sites. Coupling constans $ J_{K,H} $
        are introduced in Eq.(\ref{KCh}).
        \label{KH-lattice}
           }
\end{figure}

The goal of this paper is to demonstrate that this uncertainty can be removed by realizing CLS
in Kondo-Heisenberg systems (KHS) \cite{Doniach,Zahar-2001,KH-Lee,Tsv2016,KH-2016} which consist
of localized spins and itinerant electrons. Their coexistence leads to a competition between the
direct Heisenberg spin exchange and the
RKKY interaction generated by the electrons, see Fig.\ref{KH-lattice}. The chirality is not
explicitly broken in KHS. Therefore, if the RKKY interaction dominates, the chiral order is absent.
However, when the Heisenberg interaction exceeds some critical value, see Eqs.(\ref{condition},\ref{Jcrit})
below, one comes across an Ising-type phase transition accompanied by spontaneously breaking the chirality
and by a formation of the CSL order. This is our main result.

\begin{figure}[t]
   \includegraphics[width=0.42 \textwidth]{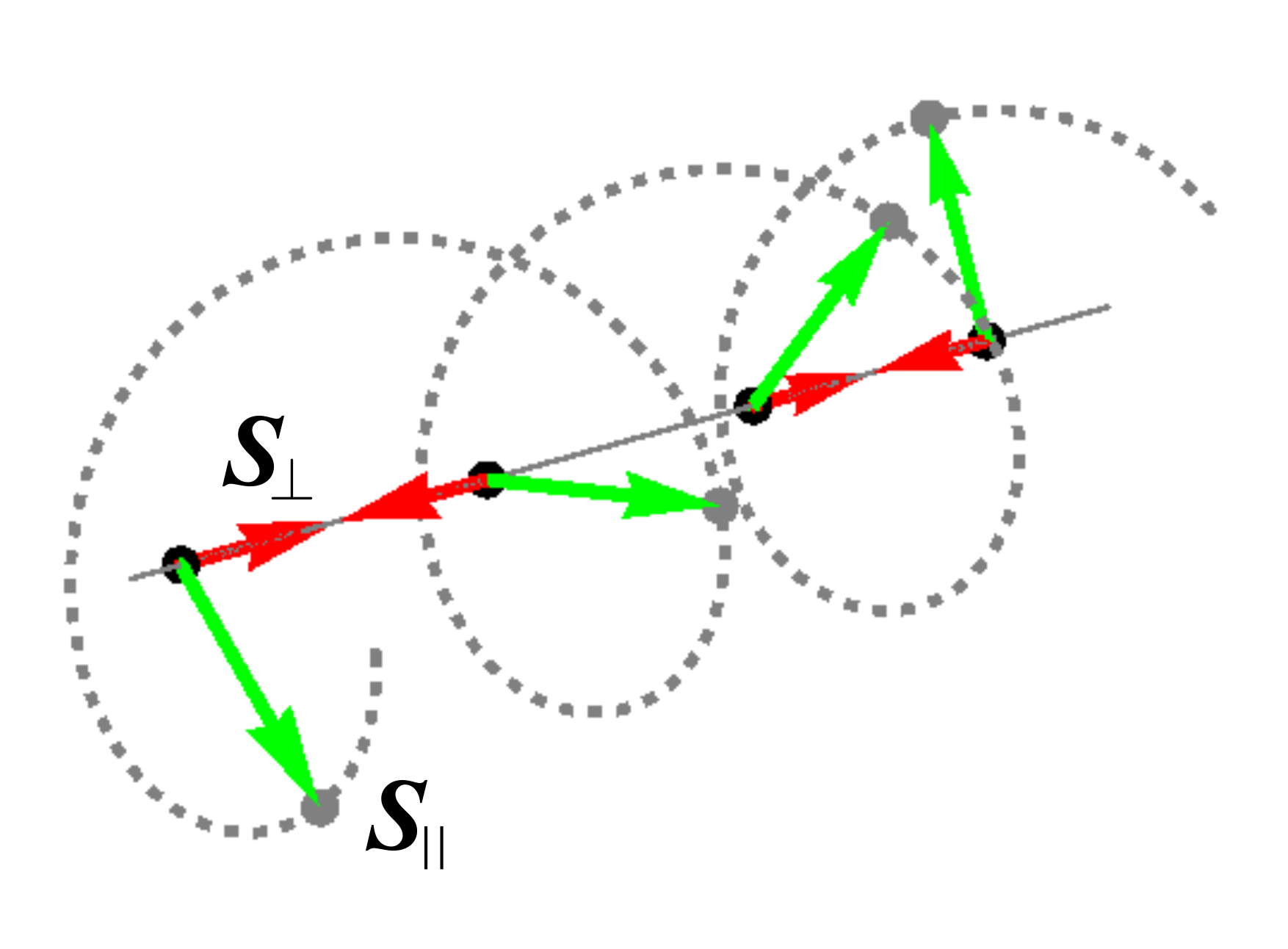}
   \caption{
        (color on-line) Chiral configuration of spins in the antiferromagnetic CSL phase. The dotted line is the helix. 
        The green and red arrows show helical-, $ \vec{S}_\parallel $,
        and antiferromagnetic, $ \vec{S}_\perp $, components of the spin, respectively, see Eq.(\ref{spins}).
        For simplicity, we disregard deformations of the helix on the scale of several
        lattice constants which is caused by the thermal fluctuations  of the triad $ \vec{e}_{1,2,3} $.
        \label{Helix-CSL}
           }
\end{figure}

We emphasize that the scalar chirality is necessary for the quantum effects mentioned above but
it does not require them and can exist in spin systems where the magnetic order is destroyed not by 
quantum, but by thermal fluctuations. We shall demonstrate that the CSL state can emerge in classical 
(quasi) two-dimensional (2D) systems when the spin susceptibility of the electron gas has a
sharp maximum at some non-zero wave vector $ \vec{Q}$ incommensurate with the lattice. 
The easiest way to model this is to assume that
the Fermi surface has nested portions. In the second order in the spin-electron coupling
constant, the Fourier transform of the RKKY exchange is proportional to the  spin susceptibility of itinerant electrons
and, hence, is strongly enhanced at $ \vec{Q}$. Without loss of generality, we can consider KHS with the spins
situated on  a 2D lattice with a short range antiferromagnetic Heisenberg exchange. The spins
interact with electrons with a nested Fermi surface. Thermal fluctuations in 2D prevent long range spin order  in
SU(2) symmetric system, but do not prevent the chiral one. When the Heisenberg exchange overwhelms the RKKY
interaction the scalar chiral order (SCO) emerges (cf. Fig.\ref{Helix-CSL}) as {\it the only non-trivial order parameter}:
\bea
\label{ChOP}
&& {\cal O}_c = \Big(\vec{S}(\vec{r}_1),[\vec{S}(\vec{r}_2)\times\vec{S}(\vec{r}_3)]\Big) .
\eea
Here, $ \vec{S} $ are the spin operators located on neighboring lattice sites $ \vec{r}_{1,2,3} $.
The energetically favorable spin configuration is presented below in Eq.(\ref{spins}).
We predict that $ {\cal O}_c $ acquires a non-zero expectation value below a certain temperature 
breaking parity and time-reversal symmetries. 
Unlike noncollinear magnets, which have other order parameters (e.g., linear in spins), the 
thermodynamic CSL phase is fully characterized by $ {\cal O}_c $. 

We will now explain how to justify our predictions.
We consider the model  combining the Kondo lattice Hamiltonian and the
Heisenberg interaction between the local moments, $ \hat{H} = \hat{H}_{K} + \hat{H}_H $, where
\bea
 \hat H_K & = & \sum_\vec{k} \epsilon(\vec{k})\hat{c}^\dagger(\vec{k})\hat{c}(\vec{k}) +
                J_K \sum_{\vec{r}} \hat{c}^\dagger(\vec{r}) \vec\s \hat{c}(\vec{r}) \vec{S}(\vec{r}),  \cr
 \hat H_H & = & J_H \sum_{\vec{r},\vec{a}} \vec{S}(\vec{r} + \vec{a}) \vec{S}(\vec{r}), \
                \vec{S} = \{S_x, S_y, S_z \}.
\label{KCh}
\eea
Here $ \hat{c}^{\rm T} \equiv ({c}_{\uparrow}(\vec{r}), c_{\downarrow}(\vec{r})) \, $ are electron
operators at lattice site $ \, \vec{r} $; $ \hat{c}(\vec{k}) $ is Fourier-transformed $ \hat{c}(\vec{r}) $;
$ \, \vec{\sigma} = \{\sigma_x, \sigma_y, \sigma_z \} \, $ are Pauli matrices; $ \, S_{x,y,z}(\vec{r})$
are components of the spin-$s$ operator $ \vec{S} $ located on lattice site $ \, \vec{r} $; $ J_{K,H} $ are
coupling constants of the isotropic exchange interaction which are much smaller than the bandwidth, $ \, s J_K, s J_H
\ll D $. The Heisenberg exchange acts between nearest neighbors, i.e., $ \vec{a} $ are smallest vectors of the lattice. 
To model the above discussed maximum of the electron spin susceptibility, we assume
that the dispersion $ \epsilon(\vec{k}) $ is nested with a wave vector $ \vec{Q}$ being
incommensurate with the lattice: $ \epsilon(\vec{k}) = - \epsilon(\vec{k} + \vec{Q})$. 
We emphasize that this is just a simple model providing the susceptibility maximum
and nesting should not be considered as a strict requirement for our theory.
The electron band is far from half filling. We concentrate on the regime where the RKKY
interactions suppresses the Kondo screening such that the latter can be neglected,
see Ref.\cite{Schimmel2016} for details. For the sake of simplicity, we will not
distinguish the crystalline lattice and the lattice of the spins. To simplify the 
calculations, we choose the 2D dispersion relation $  \epsilon(\vec{k}) = k_x^2 / 2 m_x -2 t_y \cos (k_y a_y) $,
see Suppl.Mat.1C, which is parametrized by the effective mass in the $ x $-direction, $ m_x $, and
by the hopping integral along the $ y $-direction, $ t_y $. Results will be simplified for the
case of a square 2D lattice with equal lattice constants $ a_x = a_y = a_0 $.

A one-dimensional (1D) Kondo chain, i.e. a 1D version of the model Eq.(\ref{KCh}) with
$J_H=0$, was studied in Refs.\cite{TsvYev2015, Schimmel2016}. It has been shown that, in the
case of densely located spins, the physics is dominated by the backscattering processes
which generate the RKKY exchange and suppress the Kondo screening. We have obtained
non-perturbative solutions for two limiting cases of the easy-axis and of the easy-plane
anisotropy of the Kondo exchange. In the latter case, the local spins assemble into a
quasi long range  vector chiral (or ``helical'') order, see the order parameter $ {\cal O}_h $ 
in Suppl.Mat.2.
The helix can be either left- or right handed. The spontaneously chosen helix orientation
breaks the helical symmetry of the conduction electrons which results in a (partial)
symmetry protection of the ideal transport.

\begin{figure}[t]
   \includegraphics[width=0.42 \textwidth]{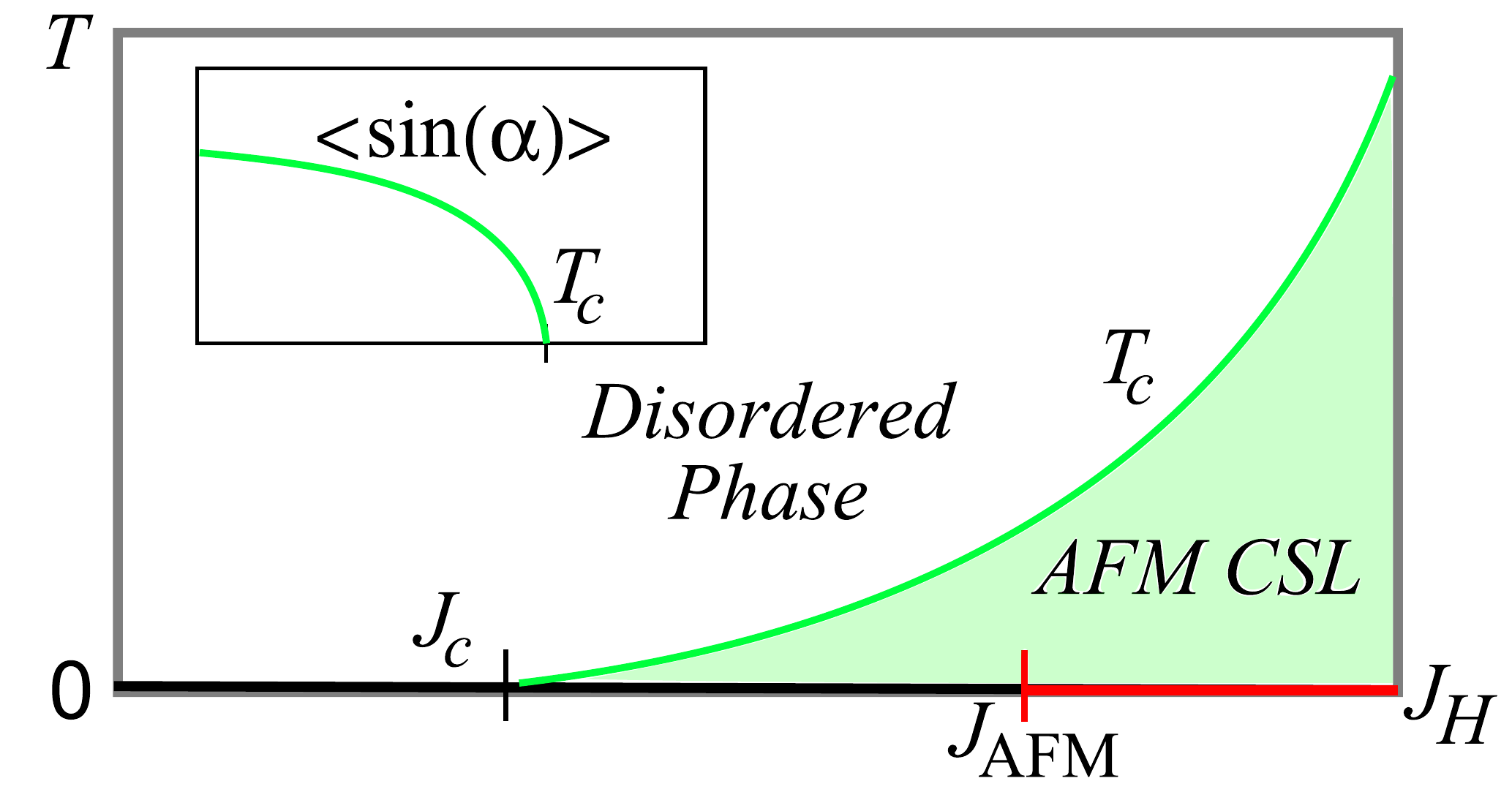}
   \caption{
        (color on-line) Phase diagram of the isotropic 2D KHS
        on the plane $ T $ vs. $ J_H $ (measured in arbitrary units) at $ T \ll s |J_K| $.
        The green line is the critical line, see Eq.(\ref{Tc-General}). It separates
        the disordered phase and CSL (green area). $ J_{\rm AFM} $
        marks the transition from CSL to the antiferromagnetic phase (red line) at $ T = 0 $.
        Inset: Temperature dependence of the mean value $ \la \sin(\alpha) \ra $. Note that
        the SCO parameter is proportional to this quantity, see Eq.(\ref{Och-nontriv}).
        \label{PhaseDiagr}
           }
\end{figure}

In this paper, we concentrate on magnetic properties of KHS.
Due to  thermal fluctuations, the helical spin ordering does not occur when the SU(2) symmetry is not broken. As we shall see,
KHS is in a disordered phase at $ J_H < J_c \sim (J_K^2 / D )  \log(D/|J_K|) $.
When $ J_H $ exceeds $ J_c $, a phase transition of the Ising-type occurs and
the spins form a (local) SCO, see
a scheme of the phase diagram on Fig.\ref{PhaseDiagr}.

To establish the existence of the CSL  it suffices to calculate the ground state energy
of our model in the proper spin background. These calculations are similar to those for the 1D Kondo chain
\cite{TsvYev2015, Schimmel2016} and, therefore, we outline them for KHS
without repeating algebraic details. Firstly, we change from the Hamiltonian to the
action and single out slow fermionic modes located at  the right- and left sheets of the
open Fermi surface, see Suppl.Mat.1C, with an ultimate aim  to develop an effective
low-energy field theory for the spins. To do this, we need to separate the  fast- and the slow
spin degrees of freedom which can be conveniently done with the help of the parametrization
\bea
\label{spins}
  \vec{S}(\vec{r})  & = & \vec{S}_\parallel(\vec{r}) + \vec{S}_\perp (\vec{r}) ; \\
  \vec{S}_\parallel & = & s \cos(\alpha) \Bigl[ \vec{e}_1 \cos(\vec{Q} \vec{r}) + \vec{e}_2 \sin(\vec{Q} \vec{r}) \Bigr]; \cr
  \vec{S}_\perp & = & s \sin(\alpha)\, (-1)^{N(\vec{r})}  \vec{e}_3; \ \  (\vec{e}_i \vec{e}_j) = \delta_{ij}.
\nonumber
\eea
The triad of mutually orthogonal unit vectors $ \vec{e}_{1,2,3} $ and the angle $\alpha$ depend
on the coordinate $ \vec{r} $ and change slowly over the lattice distance $ a_0 $. To
be definite, we choose the antiferromagnetic Heisenberg exchange on a bipartite lattice such that $N(\vec{r})$ is a sum
of all lattice coordinates for a given site.

We are interested in the state where $ \sin(\alpha) $ acquires a nonzero average below some transition
temperature and the triad of vectors $ \vec{e}_{1,2,3} $ remain disordered, at least at finite temperatures.
As we shall see, the fluctuations of  angle $ \alpha $ always remain  massive.
Its mean value will be found from minimizing the free energy.

To calculate  the  ground state energy, we first neglect space variations of the $\vec{e}_i$ vector
fields and integrate out the electrons, see Suppl.Mat.1A and 1C. We will comment on the space
variations during the second step when we derive the Landau free energy for the fluctuations. The spin
configuration (\ref{spins}) gaps out only half of the electronic modes and another half remains gapless.
A similar effect has been predicted by us for the 1D Kondo lattice where the anisotropy is of the easy plane
type and one helical sector of the fermions is gapped \cite{TsvYev2015, Schimmel2016}. However, in the
SU(2)-symmetric system, the axis of the spin spiral fluctuates in space which does not allow a global
identification of gapped and gapless fermionic modes. The density of the ground state energy for the
uniform and static configuration reads as
\bea
\label{pot}
   E_0/s^2 & = & J_H \sum_{\vec{a}} (-1)^{N(\vec{a})} \sin^2(\alpha) + \\
              + &    & \!\!\!\!\! \cos^2(\alpha)
                       \Big[
                   \tilde{J}_H(\vec{Q}) - \rho(\epsilon_F) J_K^2 \ln \Big( D/|s J_K\cos(\alpha)| \Big)
                       \Big];
\nonumber
\eea
where $\tilde J_H(\vec{q}) = J_H \sum_{\vec{a}} \cos (\vec{q} \vec{a} ) $ is the Fourier transform of the
Heisenberg  exchange
interaction; $ \rho(\epsilon_F) $ is the density of states (per one unit cell of the lattice) at the Fermi energy. 
We emphasize that, if the Fermi surface is nested, the specific choice of the dispersion relation 
has an influence only on $ \rho(\epsilon_F) $ but neither the structure of Eq.(\ref{pot}) nor its further analysis 
depend on details of $ \epsilon (\vec{k}) $.
In the case of a square 2D lattice, we obtain $  (-1)^{N(\vec{a})} = -1 $ such that $ J_H \sum_{\vec{a}}
(-1)^{N(\vec{a})} $ simplifies to $ \tilde{J}_H(\vec{G}) $ with $ \vec{G} = \{ \pi / a,  \pi / a \} $. We will use
the contracted notation $ \tilde{J}_H(\vec{G}) $ for $ J_H \sum_{\vec{a}} (-1)^{N(\vec{a})} $
below implying that $ \tilde{J}_H(\vec{G}) < 0 $.

$ E_0(\alpha) $ has three extrema: one  at $ \alpha = 0 $ and the other two at $ \alpha $ defined by
the following equation:
\be
  \label{condition}
  |\cos\alpha| = {\cal C}(J_H) \equiv
                 \frac{e^{ -\frac{1}{2} } D }{s|J_K|} \exp \left[
                                     \frac{ \tilde J_H(\vec{G}) - \tilde J_H(\vec{Q}) }{ \rho(\epsilon_F)J_K^2 }
                                                    \right].
\ee
The fluctuations of  $ \alpha $ are massive in both cases.
Since $ |\cos(\alpha)| \le 1 $, the nontrivial minimum defined in Eq.(\ref{condition}) appears only
at sufficiently strong $ J_H $. The critical value can be found from the equation
\be
  \label{Jcrit}
  {\cal C}(J_c) = 1 \ \Rightarrow \ J_c \sim \rho(\epsilon_F) J_K^2 \log(D/s|J_K|).
\ee
If $ J_H < J_c $ the minimum of the energy is located at $ \alpha = 0 $ and the system is in {\it the disordered
phase} with $ {\cal O}_c = 0 $, see Suppl.Mat.2. When $ J_H > J_c $,  the effective potential Eq.(\ref{pot})
has two equivalent minima corresponding to different signs of $\alpha \neq 0$ and two signs of the finite SCO
parameter, see Eq.(\ref{Och-nontriv}). This corresponds to {\it the CSL phase}. Since the vacuum is doubly
degenerate, the SCO parameter at $ T=0 $ reflects broken Z$_2$ symmetry and there is an Ising like phase
transition at finite temperature $T_c$. We can estimate $ T_c $ by the height of the potential barrier in the
effective potential Eq.(\ref{pot}):
\be
\label{Tc-General}
  J_H > J_c: \quad T_c \sim E_0 \bigl|_{\cos(\alpha)=1} - E_0 \bigl|_{\cos(\alpha) = \la\cos(\alpha)\ra} .
\ee
For $ J_H $ close to $ J_c $, Eq.(\ref{Tc-General}) simplifies to:
\be
\label{Tc}
   T_c \sim \rho^{-1}(\epsilon_F)
            \left[
               ( J_H - J_c ) / J_K
            \right]^2 \!\! .
\ee
At $ T < T_c(J_H) $ and $ J_H > J_c $, the SCO parameter acquires the finite value
(see Suppl.Mat.2):
\begin{widetext}
\be
\label{Och-nontriv}
    {\cal O}_{c} = s^3 \langle \sin[\alpha(\vec{r})] \cos[\alpha(\vec{r})]^2 \rangle
                       \Bigl[
                          (-1)^{N(\vec{r}_3)} \sin(\Delta_{12})
                        + (-1)^{N(\vec{r}_1)} \sin(\Delta_{23})
                        + (-1)^{N(\vec{r}_2)} \sin(\Delta_{31})
                       \Bigr] ; \
    \Delta_{jj'} \equiv (\vec{Q}, \vec{r}_j -  \vec{r}_{j'}) .
\ee
\end{widetext}

To describe fluctuations of the vector fields, we have to integrate over the fermions
and to make a usual gradient expansion keeping only leading terms, see Suppl.Mat.1.
This yields the Landau free energy density for the disordered and for the chiral phases.
At low temperatures and on the square 2D lattice we obtain:
\bea
\label{sigmaM}
 {\cal F} & = &
                     \frac{1}{8}
                        \sum_{j=1,2,3} \ \sum_{\nu = x,y} {\cal R}_{j,\nu} ( \p_{\nu}\vec{e}_j )^2 ; \
                    ({\vec e}_i, {\vec e}_j) = \delta_{ij}; \\
  {\cal R}_{1,\nu} & = & \rho(\epsilon_F) v_x^2 \, \delta_{\nu,x} -
                                    2 \la \cos^2(\alpha) \ra (s a_0)^2 J_H \cos(\vec{Q}\vec{a}_\nu),
      \cr
  {\cal R}_{2,\nu} & = & {\cal R}_{1,\nu},
      \cr
  {\cal R}_{3,\nu} & = & \rho(\epsilon_F) v_x^2 \, \delta_{\nu,x} -
                                    2 \la \sin^2(\alpha) \ra ( s a_0)^2 \tilde{J}_H(\vec{G}). \nonumber
\eea
Here $ v_{x} $ is the $x$-projection of the Fermi velocity. The stiffness tensor ${\cal R}_{j,\nu}$ in
Eq.(\ref{sigmaM}) is generically anisotropic. Its anisotropy is not universal and depends, in particular,
on a specific choice of $ \epsilon(\vec{k}) $ and on temperature.

Eq.(\ref{sigmaM}) has a form of a nonlinear sigma model with the symmetry SU(2)$\times$U(1). Sigma
models similar to Eq.(\ref{sigmaM}) have been studied in the context of noncollinear
antiferromagnetism \cite{Azaria92,Azaria93,Chubukov94}. Nonlinearity of the theory Eq.(\ref{sigmaM})
comes from the orthonormality of the vectors $ {\vec e}_j $. In 2D, this interaction
generates a finite  correlation length $ \xi $ \cite{ATsBook}.
In the renormalization procedure, this manifests itself as a continuous decrease of components
of the stiffness ${\cal R}_{j,\nu}(\Lambda)$ with the decrease of the momentum cut-off $\Lambda$.
As a result, the fluctuations acquire a correlation length which is exponentially large in
${\cal E}_{\rm UV}/T$; $ {\cal E}_{\rm UV} $ is the UV regularizer, cf. Refs.\cite{2D-AFM-1,2D-AFM-2}.

We consider the finite temperatures implying that thermal
fluctuations dominate over the quantum ones at length scales $ L > \xi > v/T $, where $ v $ is a
characteristic velocity of the spin excitations. In this case, one can treat the fields $ \vec{e}_i $
as time independent and there is no need to promote the free energy description to the full
dynamical theory. The thermal fluctuations prevent a breaking of the SU(2) symmetry of
Eq.(\ref{sigmaM}) and the magnetic order can occur only at zero temperature, see Fig.\ref{PhaseDiagr}.
Thus, at $ J_H > J_c $ and $ T \ne 0 $ this leaves us with SCO as the only possible order.

One has to distinguish two regimes where the theory Eq.(\ref{sigmaM}) can be used: 1) The model with
$\alpha =0$
corresponds to the disordered phase and can be used in the
temperature interval between the Ising transition temperature and the fermionic gap: $ T_c \ll T \ll sJ_K $.
2)~The model with $\alpha \ne 0$
corresponds to CSL and should be used well below the Ising transition,
$ T_{\rm min} < T \ll T_c $, where one can neglect fluctuations of $\la\sin\alpha\ra $.

Although all quantum effects in CSL are very interesting we leave their systematic study
for the forthcoming paper. At present, we can make only a preliminary guess: We note that the charge 
and the spin degrees of freedom are deeply connected in our approach, see Suppl.Mat.3. The
Kondo lattice model considered in Refs.\cite{TsvYev2015, Schimmel2016} has the same property.
Based on this analogy and on the fully quantum theory of Refs.\cite{TsvYev2015, Schimmel2016},
we surmise that nontrivial excitation of the KHS are slow spinons dressed by localized electrons.

{\it To summarize}, we have found that increasing the direct Heisenberg exchange in the Kondo-Heisenberg
model with the nested Fermi surface leads to a phase transition to the state with  spontaneously broken
scalar chirality. The corresponding chiral (local) order parameter, $ {\cal O}_c $ in Eq.(\ref{ChOP}), 
breaks time reversal and parity symmetry. This symmetry is  Z$_2$ and  the transition belongs to the universality
class of the Ising model.

We believe that KHS can be used as a principally new platform to realize SCO  
in non-exotic experimental setups. Our finding paves the way towards removing the doubt whether the 
chiral spin liquid with the scalar chirality
can exist in the realistic systems where the time-reversal symmetry is not explicitly broken.

The broken time reversal and parity symmetry can reveal itself
in the optical measurements through, for instance, the Kerr effect or measurements of nonlinear optical responses.
The second harmonic response is particularly sensitive to the presence of global inversion symmetry. There are two other,
though not definite, experimentally detectable indicators which can complement the optical experiments
and confirm formation of CSL, namely, peculiar magnetic- and electronic responses of the
antiferromagnetic KHS with the nested Fermi surface. Firstly, the energetically favorable spin configuration,
Eq.(\ref{spins}), suggests that correlation functions of all spin components have $ \vec{Q} $-harmonics.
Therefore, spin susceptibilities possess the Bragg peaks not only on the Neel vector but also on the wave
vectors $ \pm \vec{Q}$. These new peaks are smeared out by smooth fluctuations of the spin
$ \vec{Q} $-components, including the fluctuations of the triad $ \vec{e}_{1,2,3} $ and of the angle
$ \alpha $. The triad fluctuations are (almost) insensitive to the Ising phase transition at  $ J_H > J_c, \,
T \to T_c $. However, the fluctuations of $ \alpha $ are suppressed in the CSL phase and, therefore,
the peaks become sharper at $ J_H > J_c, \, T < T_c $. On the other hand, the response of the itinerant
electrons will experience a drop when the probe frequency and the temperature are  below $sJ_K >T_c$.
Such a drop is related to the fact that one half of the electrons acquire  a gap while the other half remains
gapless. This decrease in the number of carriers is expected to alter the electric properties of
a sample, cf. Ref.\cite{HeliStat-GaAs}; more details will be presented elsewhere.

A model, which is described by the Kondo part of our Hamiltonian, $ \hat{H} = \hat{H}_K $ at $ J_H=0 $,  has 
been considered in Ref.\cite{martin_itinerant_2008} on  triangular lattice. It has been demonstrated that, for a 
particular band filling providing two independent nesting vectors of the Fermi surface, the chiral order is formed. We 
would like to stress that our approach is much more general and does not require any special fine tuning.
Particularly, details of the band dispersion are not important 
for our general predictions. The only crucial ingredient is the strong maximum of the spin susceptibility of the 
itinerant electrons. A nested Fermi surface is just a simple way to achieve it and should not be considered 
as a strict requirement for our theory imposing restrictions on its experimental verification.
Possible candidates for the experimental realization of the described Kondo-Heisenberg system
with the spontaneously broken chirality are proximity-coupled layers of metals and Mott-insulators.
At present, we know at least one system which is structurally similar to what we propose. This is
Sr$_2$VO$_3$FeAs, a naturally assembled heterostructure made of well separated layers of an
iron-based metal SrFeAs and Mott-insulating vanadium oxide.

\begin{acknowledgments}
{\bf Acknowledgments}:
A.M.T. was supported by the U.S. Department of Energy (DOE), Division of Materials Science,
under Contract No. DE-AC02-98CH10886. O.M.Ye. acknowledges support from the DFG through the
grant YE 157/2-1.
We are grateful to Igor Mazin, Jan von Delft, Matthias Punk,
and Denys Makarov for useful discussions.
\end{acknowledgments}

\bibliography{Bibliography}

\widetext

\newpage

\appendix

\section{Supplementary material}

\subsection{1. Derivation of the Landau Functional}

\subsubsection{1.A The case of a 1D system at $ J_H = 0 $}

Let us for the moment neglect the direct exchange and consider a purely 1D system where $ Q = 2 k_F $; $ k_F $
is the Fermi momentum. At first, we parameterize the Kondo interaction term with a $ 2 \times 2 $ SU(2) matrix with
$ 2 k_F $-component $ g $, select the most relevant backscattering terms (cf. Refs.\cite{TsvYev2015, Schimmel2016}):
\bea
V_K = \bar{J}(R^+ g \s_- g^{-1} L + H.c.); \quad \bar{J}= s J_K \cos(\alpha);
\eea
and rotate the fermions $ g^{-1}L \rightarrow L', \quad g^{-1}R \rightarrow R'$. This rotation is anomaly free.
In the rotated basis, the inverse Green function of the fermions can be written as $ \hat{G}^{-1}
= \hat{G}_0^{-1} + \hat{\Omega} $ with
\be
  \hat{G}_0^{-1} = \left(
            \begin{array}{cccc}
                \partial_+ &      0     &     0      &    0    \\
                    0      & \partial_+ & \bar{J}  &    0    \\
                    0      & \bar{J}  & \partial_- &    0    \\
                    0      &      0     &     0      & \partial_-
            \end{array}
                    \right); \quad
  \hat{\Omega} = \left(
            \begin{array}{cccc}
                \Omega^z_+ &   \Omega^-_+ &     0      &    0      \\
                \Omega^+_+ & - \Omega^z_+ &     0      &    0      \\
                    0      &      0       & \Omega^z_- & \Omega^-_-\\
                    0      &      0       & \Omega^+_- & - \Omega^z_-
            \end{array}
                 \right) ; \quad
       \Omega^a_{\pm} \equiv \frac{1}{2}\mbox{tr}[\s^a g^{-1}(\p_{\tau} \pm \ri v_F \p_x)g].
\ee
$ \hat{G}_0 $ describes the fermions in the case of the homogeneous and static spin configuration
and $ \hat{\Omega} $ reflects an influence of the spin fluctuations. It is important for further
calculations that the constant spin configuration induces the energy gap in one sector of the
rotated fermions; the second sector remains gapless. These two sectors are coupled only by the
spin fluctuations.

We are interested in properties of the spin subsystem. Therefore, one can
integrate out {\it all} fermions, exponentiate the fermionic determinant as $ \rm{tr} \log [\hat{G}_0^{-1}
+ \hat{\Omega}]$ and expand it in $ \hat{\Omega} $ up to quadratic terms. We note in passing that $ \rm{tr}
\log [\hat{G}_0^{-1} ] $ determines the ground states energy and linear terms $ O(\hat{\Omega}) $ are absent
in the expansion. The calculation of the ground state energy is described in details in Refs.\cite{TsvYev2015, Schimmel2016},
we do not repeat it here. The expansion in fluctuations yields a sum of responses $ \Omega^a_\mu(\Omega,P)
\Pi_{AB} \Omega^{a'}_{\mu'}(-\Omega,-P) $ with $ (\Omega,P) $ being small frequency and momentum of gapless
fluctuations of the spin configuration. We will skip below the external frequency and momentum and mark
vertices with the inverted argument by bars. The fermionic contribution to the Lagrangian of the spins subsystem is:
\be
\label{Resp}
  \delta {\cal L} =
  - \frac{1}{2} \left\{
      \Omega^z_+ ( \Pi_{RR} + \Pi_{F^-F^-} ) \bar{\Omega}^{z}_{+} +
      \Omega^z_- ( \Pi_{LL} + \Pi_{F^+F^+} ) \bar{\Omega}^{z}_{-} 
      -2 \Omega^z_+ \Pi_{BB} \bar{\Omega}^{z}_{-} +
       2 \Omega^-_- \Pi_{LF^+} \bar{\Omega}^{+}_{-} +
       2 \Omega^+_+ \Pi_{RF^-} \bar{\Omega}^{-}_{+}
               \right\} \!.
\ee
The response function read as
\be
  \Pi_{AB} = \int_{-\infty}^{\infty} \frac{{\rm d} \, k}{2 \pi} \
        \left[ T \sum_{\omega_n} G_A G_B \right];
\ee
(mind the order of summations)
where $ k $ and $ \omega_n = (2 n + 1) \pi T $ are the internal momentum and the (fermionic
Matsubra) frequency, respectively, and we have introduced the Green functions of the gapless
fermions
\be
  G_{R/L} = \frac{1}{i \omega_m \mp v_F k}
\ee
and of the gapped fermions:
\be
  G_{F^{\mp}} = - \frac{i \omega_n \pm v_F k}{\omega_n^2 + (v_F k)^2 + \bar{J}^2}; \
  G_B = \frac{\bar{J}}{\omega_n^2 + (v_F k)^2 + \bar{J}^2} \, .
\ee
Next we note that the small external energy and momentum can be neglected in all
$ \Pi_{AB} $ containing at least one Green function of the gapped fermions. Calculating all
Matsubara sums and momentum integrals, we obtain
\bea
      \Pi_{RR/LL} & = & 
\rho_{1D} \frac{\pm v_F P} {i \Omega \mp v_F P} ; \quad 
      \Pi_{BB} \simeq \rho_{1D} \left\{
                        \begin{array}{l}
                            1/2, \ T/\bar{J} \ll 1; \\
                             O\bigl( [\bar{J}/T]^2 \bigr), \ T/\bar{J} \gg 1;
                        \end{array}
                      \right. \cr
                      \nonumber \\
      \Pi_{LF^+} & = & \Pi_{RF^-} = \Pi_{F^-F^-} = \Pi_{F^+F^+} \simeq - \rho_{1D}
                    \left\{
                        \begin{array}{l}
                           1/2, \ T/\bar{J} \ll 1; \\
                           1,   \ T/\bar{J} \gg 1.
                        \end{array}
                    \right.
      \nonumber
\eea
Here $ \rho_{1D} = 1/2 \pi v_F $ is the DoS of the 1D Dirac fermions. Inserting expressions of the
response functions into the equation for $ \delta {\cal L} $, we arrive at the rigidity of the spin
waves. In the static (classical) limit, $ \Omega \to 0 $, and at low temperatures it reduces to:
\be
\label{Rigidity}
T \ll \bar{J}: \quad
  \delta {\cal L} =
      \rho_{1D}  \left\{
      \Omega^x_P \Omega^x_{-P} + \Omega^y_P \Omega^y_{-P} + \Omega^z_P \Omega^z_{-P}
                                       \right\};
\ee
Here we have substituted $ \pm \Omega^a_P \Omega^b_{-P} $ for $ \Omega^a_\mu \bar{\Omega}^b_{\pm \mu} $.

\subsubsection{1.B Parametrization of the fluctuations by a unit vector}

Using the matrix identities
\be
\label{MatrId}
 \left\{
 \begin{array}{l}
  \hat{A} = A^{(j)} \sigma_j, \quad  A^{(j)} = \frac{1}{2} {\rm tr}[\sigma_j \hat{A}]; \\ \\
  {\rm tr}[ \vec{\sigma} \hat{A}^{-1} \sigma_j \hat{A}] \, {\rm tr}[ \vec{\sigma} \hat{A}^{-1} \sigma_{j'} \hat{A}] =
        4 \delta_{j,j'}
 \end{array}
 \right. \qquad
 j, j' = x, y, z.
\ee
one can express the interaction term with the help of the unit vector
\be
  g \sigma_- g^{-1} = \frac{1}{\sqrt{2}} (\vec{E}, \vec{\sigma}), \quad
  \vec{E} \equiv \frac{1}{\sqrt{2}} {\rm tr}[ \vec{\sigma} g \sigma_- g^{-1}];
\ee
and some (real) orthogonal basis $ \vec{e}_{1,2,3} $, for example:
\be
\label{BasisFromSU2}
   \vec{e}_{1,2,3} = \frac{1}{2} {\rm tr}[ \vec{\sigma} g \sigma_{x,y,z} g^{-1}] .
\ee
The vector conjugated to $ \vec{E} $ is $ \vec{E}^* = (1/\sqrt{2})
{\rm tr}[ \vec{\sigma} g \sigma_+ g^{-1}] $. Using Eq.(\ref{MatrId}), it
is straightforward to check orthonormality of the basis $ \vec{e}_{1,2,3} $
and equalities
\be
\label{EvecProp}
  (\vec{E}, \vec{E}) = (\vec{E}^*, \vec{E}^*) = 0; \
  || \vec{E} || = (\vec{E}, \vec{E}^*) = 1.
\ee
$ \vec{E} $ can be expanded in the basis $ \vec{e}_{1,2,3} $ as follows:
\be
\label{E-expand}
   \vec{E}   =   \frac{1}{\sqrt{2}} \left( \vec{e}_{1} - i \vec{e}_{2} \right) \, ,
            \quad
   \vec{E}^* =   \frac{1}{\sqrt{2}} \left( \vec{e}_{1} + i \vec{e}_{2} \right) \, .
\ee
Eq.(\ref{E-expand}) agrees with singling out smooth modes (without $ 2 Q$-oscillations) with the
help of Eq.(\ref{spins}).

Let us introduce gradients of the vector field which are expanded in terms of the basis $ \vec{e}_{1,2,3} $:
\be
   \p_{\mu}\vec{e}_a = \epsilon^{abc}\vec{e}_b {\cal E}_{\mu,c}, \quad
   {\cal E}_{\mu,a}=  \epsilon^{abc}(\vec{e}_b, \p_{\mu}\vec{e}_c); \quad a, b, c = 1 \ldots 3.
\ee
It is easy to prove that
\be
  ( \p_{\mu}\vec{e}_a )^2 = \sum_{b \ne a}( {\cal E}_{\mu,b} )^2 \, \quad \Rightarrow \quad
  \sum_{a=1}^3 ( \p_{\mu}\vec{e}_a )^2 = 2 \sum_{b=1}^3( {\cal E}_{\mu,b} )^2.
\ee
A straightforward algebra yields the relation between the gradients $ \Omega_\mu $ and
$ {\cal E}_{\mu} $:
\be
\label{GradsRel}
  \left\{
  \begin{array}{l}
   {\cal E}_{\mu,1} = - 2 i \Omega^x_\mu \, , \cr
   {\cal E}_{\mu,2} = - 2 i \Omega^y_\mu \, , \cr
   {\cal E}_{\mu,3} = - 2 i \Omega^z_\mu \, ,
  \end{array}
  \right.
   \quad \Rightarrow \quad
   \sum_{j=x,y,z} (\Omega^j_\mu)^2 =
        - \frac{1}{4} \sum_{a=1}^3( {\cal E}_{\mu,a} )^2 =
        - \frac{1}{8} \sum_{a=1}^3 ( \p_{\mu}\vec{e}_a )^2 .
\ee
Hence, we can rewrite Eq.(\ref{Rigidity}) as follows:
\be
\label{Rigidity-e}
  \delta {\cal L} =
       \frac{\rho_{1D} v_F^2}{8}  \sum_{a=1}^3 ( \p_{x}\vec{e}_a )^2 .
\ee

\subsubsection{1.C The case of a 2D system at $ J_H = 0 $}

As an example of a spectrum with a nested Fermi surface we will choose the following 2D dispersion:
\bea
 \epsilon(k) = \frac{k_x^2}{2 m_x} -2 t_y \cos (k_y a_y)
\eea
The Fermi surface is open and, thus, the electrons can be divided into ``left'' and ``right''
species in the vicinity of the Fermi energy.  The kinetic energy of the electrons near the Fermi surface is
\bea
H_{kin} = \left( R^\dagger_\uparrow, L^\dagger_\uparrow \right)
                            \left(
\begin{array}{cc}
[v_F k_x - 2 t_y \cos(k_y a_y)] &               0                           \\
               0              & -[v_F k_x - 2 t_y \cos(k_y a_y)]
\end{array}
                            \right)
             \left(
               \begin{array}{l}
                  R_\uparrow \\
                  L_\uparrow
               \end{array}
             \right)
          + \{ R/L_\uparrow \to R/L_\downarrow \} ;
\eea
with the nesting vector being $ \vec{Q} = (2 k_F, \pi / a_y)$.
Calculation of the Landau functional for the 2D system is very similar to that of the 1D one.
The expression for the gradients reads now:
\be
  {\rm 2D}: \qquad
  \Omega^a_{\pm} = \frac{1}{2}\mbox{tr} \Bigl\{ \s^a g^{-1}
                \left[ \p_{\tau}
                    \pm \left( \ri v \p_x + t_y a_y^2 \p_y^2 \right)
                \right] g
                                             \Bigr\} + O\left(\p_y^3 g\right) \, .
\ee
Higher gradients can be neglected while terms $ \propto g^{-1} \p_y^2 g $ do not contribute
to the rigidity [they generate Hartree-like diagrams with zero energy and momentum and,
therefore, vanish]. The second order terms reduce to the same sum of the responses as in the
1D case, see Eq.(\ref{Resp}), where the response functions acquire ad additional moment integral:
\be
  {\rm 2D}: \qquad
    \Pi_{AB} = \int_{-\infty}^{\infty} \frac{{\rm d}^2 \, k_{x,y}}{(2 \pi)^2} \
        \left[ T \sum_{\omega_n} G_A G_B \right] .
\ee
The integral over $ k_y $ is calculated trivially: one should change the momentum variable
$ k_x - 2 (t_y/v_F) \cos(k_y a_y) \to k_x' $. After this transformation, the 2D response
functions are reduce to their 1D counterparts which are multiplied by the integral of the
Jacobian over $ k_y $:
\be
  P_y = \frac{2}{\pi a_y} \int_{0}^{2 t_y / v_F} \frac{{\rm d} k_y}{\sqrt{ (2 t_y / v_F)^2 - k_y^2 }}
            = a_y^{-1} .
\ee
Finally, we obtain the same answer Eq.(\ref{Rigidity}) where the 1D DoS must be changed
to the 2D one:
\be
  \rho_{2D} = \rho_{1D} P_y = \frac{1}{2 \pi v_F a_y} .
\ee
Hence, the only difference between the Landau functionals calculated for the 1D and 2D
cases is in the renormalization of the DoS.

\subsubsection{1.D Contribution of the direct exchange to the rigidity of excitations}

A contribution of the Heisenberg exchange to the rigidity of the excitations
is independent on that of the RKKY exchange and can be calculated straightforwardly in
the parametrization of the basis $ \vec{e}_{1,2,3} $, see Eq.(\ref{spins}). The smooth
part (averaged of $ 2 Q $-oscillations) of the exchange energy reads:
\bea
  {\cal L}_H & = & s^2 J_H \left(
     \frac{\cos^2(\alpha)}{2} \Bigl\{ \cos( {\vec{Q} \vec{a}} )
                    \Bigl[
             \bigl(
          \vec{e}_1 (\vec{r}+\vec{a}), \vec{e}_1(\vec{r})
             \bigr) +
             \bigl(
          \vec{e}_2 (\vec{r}+\vec{a}), \vec{e}_2(\vec{r})
             \bigr)
                     \Bigr] +
                                             \right. \cr
                     &   &
                   \left. + \, \sin( {\vec{Q} \vec{a}} )
                    \Bigl[
             \bigl(
          \vec{e}_2 (\vec{r}+\vec{a}), \vec{e}_1(\vec{r})
             \bigr) -
             \bigl(
          \vec{e}_1 (\vec{r}+\vec{a}), \vec{e}_2(\vec{r})
             \bigr)
                    \Bigl]
                    \Bigr\} +
     (-1)^{N(\vec{a})} \sin(\alpha)^2
                    \bigl( \vec{e}_3 (\vec{r}+\vec{a}), \vec{e}_3(\vec{r}) \bigr)
                    \right) \, .
\eea
In this section, we skip summations over $ \vec{r} $ and $ \vec{a} $, see Eq.(\ref{KCh}) in the main text, and
disregard an unimportant variation of $ \alpha $ on the neighboring lattice sites.
Now, we have to expand the vectors $ \vec{e}_j (\vec{r}+\vec{a}) $ in powers of $ \vec{a} $:
\[
  \vec{e}_j (\vec{r}+\vec{a}) \simeq
      \vec{e}_j (\vec{r}) + \p_{\nu} \vec{e}_j (\vec{r}) a_\nu +
      \frac{1}{2} \p^2_{\nu,\nu'} \vec{e}_j (\vec{r}) a_\nu a_{\nu'}; \ \nu, \nu' = x, y \, .
\]
The leading term yields the contribution of the Heisenberg exchange to the ground state energy
\be
  {\cal L}_H^{(0)} = s^2 J_H \left(
                \cos^2(\alpha) \cos( {\vec{Q} \vec{a}} ) + (-1)^{N(\vec{a})} \sin^2(\alpha)
                             \right) \, .
\ee
Linear terms are absent in the expansion around the minima of the energy.
The second order terms yield
\bea
\label{L2_H}
  {\cal L}_H^{(2)} = \frac{s^2}{2} & J_H & \sum_{\nu,\nu'} a_\nu a_{\nu'}
                            \Biggl(
     \frac{ \langle \cos^2(\alpha) \rangle }{ 2 }
         \Bigl\{
             \cos( {\vec{Q} \vec{a}} )
                    \Bigl[
             ( \p^2_{\nu,\nu'} \vec{e}_1 (\vec{r}), \vec{e}_1(\vec{r}) ) +
             ( \p^2_{\nu,\nu'} \vec{e}_2 (\vec{r}), \vec{e}_2(\vec{r}) )
                    \Bigr] + \\
                    &   &
          +  \, \sin( {\vec{Q} \vec{a}} )
                   \Bigl[
                       ( \p^2_{\nu,\nu'} \vec{e}_2 (\vec{r}), \vec{e}_1(\vec{r}) ) -
                       ( \p^2_{\nu,\nu'} \vec{e}_1(\vec{r}), \vec{e}_2 (\vec{r}) )
                   \Bigr]
          \Bigr\} +
     (-1)^{N(\vec{a})} \langle \sin^2(\alpha) \rangle
                    ( \p^2_{\nu,\nu'} \vec{e}_3 (\vec{r}), \vec{e}_3(\vec{r}) )
                             \Biggr)  \, .
\nonumber
\eea
After integrating by parts, Eq.(\ref{L2_H}) generates the exchange contribution to the rigidity:
\bea
  {\cal L}_H^{(2)} = - \frac{s^2}{2} & J_H & \sum_{\nu,\nu'} a_\nu a_{\nu'}
                            \Biggl(
     \frac{ \langle \cos^2(\alpha) \rangle }{ 2 }
             \cos( {\vec{Q} \vec{a}} )
                    \Bigl[
             ( \p_{\nu} \vec{e}_1 (\vec{r}), \p_{\nu'} \vec{e}_1(\vec{r}) ) +
             ( \p_{\nu} \vec{e}_2 (\vec{r}), \p_{\nu'} \vec{e}_2(\vec{r}) )
                    \Bigr] + \\
                    &   &
          +  \, 
     (-1)^{N(\vec{a})} \langle \sin^2(\alpha) \rangle
                    ( \p_{\nu} \vec{e}_3 (\vec{r}), \p_{\nu'} \vec{e}_3(\vec{r}) )
                             \Biggr)  \, .
\nonumber
\eea
Here, the classical value (obtained from the minimization of the energy) must be substituted
for the massive angle $ \alpha $.

\subsection{2. The order parameters}

The spin order in the Kondo-Heisenberg system is characterized by the helical
and the chiral order parameters:
\be
    {\cal O}_{h} =  [ \vec{S}(\vec{r}_2) \times \vec{S}(\vec{r}_1) ]  , \quad
    {\cal O}_{c} =  ( \vec{S}(\vec{r}_3), [ \vec{S}(\vec{r}_2) \times \vec{S}(\vec{r}_1) ] ) .
\ee
Below, we will assume that $ \vec{r}_{1,2,3} $ are (close to) neighboring lattice sites and
average over spin oscillations treating $\alpha$ and ${\vec e}_a$ as slowly varying fields
on the scale of the lattice spacing, i.e., we will neglect their spacial variations.

Let us use the parametrization Eq.(\ref{spins}) and single out smooth (without $ 2 Q r_j $-oscillations)
parts of $ {\cal O}_{h,c} $. The chiral order parameter has only two non-zero contributions:
\be
\label{Ohel}
  {\cal O}_{h} = \frac{s^2}{2} \langle \cos[\alpha(\vec{r}_2)] \cos[\alpha(\vec{r}_1)] \rangle \sin(\Delta_{21})
                              \left\langle
                               \bigl[ \vec{e}_2(\vec{r}_2) \times \vec{e}_1(\vec{r}_1) \bigr] -
                               \bigl[ \vec{e}_1(\vec{r}_2) \times \vec{e}_2(\vec{r}_1) \bigr]
                              \right\rangle \,; \quad \Delta_{jj'} \equiv \vec{Q} (\vec{r}_j - \vec{r}_{j'}) \, .
\ee
After neglecting the difference between $\vec{e}_a$ vectors on different sites, Eq.(\ref{Ohel}) reduces to
\be
  {\cal O}_h(\vec{r}) = - s^2 \langle
                                \cos[\alpha(\vec{r})]^2 \rangle \sin(\Delta r_{21})\langle \vec{e}_3(\vec{r})
                              \rangle.
\ee
Since $ \langle \vec{e}_3(\vec{r}) \rangle = 0 $ in the isotropic system, we find that $ {\cal O}_h = 0 $.
This is because a finite $ {\cal O}_h $ would reflect broken O(3) symmetry in the spin space and this
 symmetry cannot be spontaneously broken in low dimensional isotropic systems which we consider.

The smooth chiral order parameter is the sum of six contributions:
\bea
\label{Och}
    {\cal O}_{c} & = & \frac{s^3}{2} \langle \sin[\alpha(\vec{r}_3)] \cos[\alpha(\vec{r}_2)] \cos[\alpha(\vec{r}_1)] \rangle
                                                    \times \\
                 &   & \Bigl\langle
                    - \Bigl( \vec{e}_{1}(\vec{r}_3), (-1)^{N(\vec{r}_2)} \sin(\Delta_{31})
                                                     \bigl[ \vec{e}_3(\vec{r}_2) \times \vec{e}_2(\vec{r}_1) \bigr] +
                                                   (-1)^{N(\vec{r}_1)} \sin(\Delta_{32})
                                                     \bigl[ \vec{e}_2(\vec{r}_2) \times \vec{e}_3(\vec{r}_1) \bigr] \Bigr)
                        + \cr
                 &   & +
                    \Bigl( \vec{e}_{2}(\vec{r}_3), (-1)^{N(\vec{r}_1)} \sin(\Delta_{32})
                                                     \bigl[ \vec{e}_1(\vec{r}_2) \times \vec{e}_3(\vec{r}_1) \bigr] +
                                                   (-1)^{N(\vec{r}_2)} \sin(\Delta_{31})
                                                     \bigl[ \vec{e}_3(\vec{r}_2) \times \vec{e}_1(\vec{r}_1) \bigr] \Bigr)
                        + \cr
                 &    & +
                    (-1)^{N(\vec{r}_3)} \sin(\Delta_{21})
                    \Bigl( \vec{e}_{3}(\vec{r}_3), \bigl[ \vec{e}_2(\vec{r}_2) \times \vec{e}_1(\vec{r}_1) \bigr] -
                                                     \bigl[ \vec{e}_1(\vec{r}_2) \times \vec{e}_2(\vec{r}_1) \bigr] \Bigr)
                       \Bigr\rangle \, .
                 \nonumber
\eea
As above, we neglect differences between the spin variables located at different points in space. Then,
Eq.(\ref{Och}) becomes:
\be
\label{Och-hom}
    {\cal O}_{c} = s^3 \langle \sin[\alpha(\vec{r})] \cos[\alpha(\vec{r})]^2 \rangle
                                                    \times
                       \Bigl[
                          (-1)^{N(\vec{r}_3)} \sin(\Delta_{12})
                        + (-1)^{N(\vec{r}_1)} \sin(\Delta_{23})
                        + (-1)^{N(\vec{r}_2)} \sin(\Delta_{31})
                       \Bigr] \, .
\ee
The product $ \sin[\alpha(\vec{r})] \cos^2[\alpha(\vec{r})] $ acquires a finite average via the Ising phase transition
at $ J_H > J_c $, see the main text. The expression in the square brackets is nonzero for certain site arrangements.
For example, for three consecutive sites on a 1D chain with the lattice spacing $ a_0 $, it is equal to $(-1)^{N(x_2)}
\sin(2 k_F a_0)$.

\subsection{3. Connection between the charge and the spin degrees of freedom}

It is interesting to trace the connection between the charge and the spin degrees of freedom in our model.
To do this, we choose an explicit parametrization of the spins by the Euler angles:
\bea
   \vec{S} & = & s \Big\{
                     \cos(\alpha) \big[
                                  \vec{u}_1 \cos( \vec{Qr} + \beta ) + \vec{u}_2 \sin( \vec{Qr} + \beta )
                                  \big]  
                 + (-1)^{ N(\vec{r}) } \sin(\alpha)  \vec{u}_3
                   \Big\} ; \\
   \vec{u}_1 & = & \bigl( 0, -\sin(\psi), \cos(\psi) \bigr), \ 
   \vec{u}_2   = [ \vec{u}_1 \times \vec{u}_3], \ 
   \vec{u}_3   = \bigl( \cos(\theta), \sin(\theta) \cos(\psi), \sin(\theta) \sin(\psi) \bigr).
\nonumber
\eea
The connection is clearly visible in the linear combination of the spins angle $ \beta $ and the product
$ \vec{Qr} $ where the nesting vector of the Fermi surface $ \vec{Q}$ is related to the fermionic density.

\end{document}